\documentclass{emulateapj}
\usepackage{apjfonts}
\usepackage{epsfig}

\journalinfo{The Astrophysical Journal}
\slugcomment{Accepted for publication, 2011 Nov 24}

\shortauthors{CAMILO ET AL.}
\shorttitle{THE INTERMITTENT AND HIGHLY SCATTERED PSR~J1841--0500}

\begin{document}

\def\cxo{{\em Chandra}}

\def\psr{PSR~J1841--0500}
\def\psrb{PSR~B1931+24}
\def\psrc{PSR~J1832+0029}
\def\mag{1E~1841--045}
\def\snr{Kes~73}

\title{PSR~J1841--0500: a radio pulsar that mostly is not there }

\author{F.~Camilo\altaffilmark{1},
  S.~M.~Ransom\altaffilmark{2},
  S.~Chatterjee\altaffilmark{3},
  S.~Johnston\altaffilmark{4},
  and P.~Demorest\altaffilmark{2}
}

\altaffiltext{1}{Columbia Astrophysics Laboratory, Columbia University,
  New York, NY~10027, USA}
\altaffiltext{2}{National Radio Astronomy Observatory, Charlottesville,
  VA~22903, USA}
\altaffiltext{3}{Department of Astronomy, Cornell University, Ithaca,
  NY~14853, USA}
\altaffiltext{4}{Australia Telescope National Facility, CSIRO, Epping,
  NSW~1710, Australia}

\begin{abstract}
In a search for radio pulsations from the magnetar \mag, we have
discovered the unrelated pulsar J1841--0500, with rotation period
$P=0.9$\,s and characteristic age 0.4\,Myr.  One year after discovery
with the Parkes telescope at 3\,GHz, radio emission ceased from this
bright pulsar.  After 580 days, emission resumed as before.  The $\dot P$
during both on states is 250\% of the average in the off state.  \psr\ is
a second example of an extremely intermittent pulsar, although with a much
longer off period and larger ratio of spin-down rates than \psrb.  The new
pulsar is hugely scattered by the ISM, with a fitted timescale referenced
to 1\,GHz of $\tau_1 = 2$\,s.  Based on polarimetric observations at
5\,GHz with the Green Bank Telescope, the intrinsic pulse profile has not
obviously changed between the two on states observed so far, although
relatively small variations cannot be excluded.  The magnitude of its
rotation measure is the largest known, $\mbox{RM}=-3000$\,rad\,m$^{-2}$,
and with a dispersion measure $\mbox{DM}=532$\,pc\,cm$^{-3}$ implies a
large electron-weighted average magnetic field strength along the line
of sight, $7\,\mu$G.

\end{abstract}


\section{Introduction} \label{sec:intro} 

Rotation-powered pulsars spin down gradually, torqued by their
magnetospheric fields and currents.  While generally highly predictable,
the rotation in some of these neutron stars is occasionally punctuated
by ``glitches'', sudden period decreases caused by the irregular
transfer of angular momentum from the interior to the crust or by
crustal rearrangement.  Apart from glitches, the observed evolution
of rotation is dominated by the predictable effects of electromagnetic
torques, but for most pulsars there remains a seemingly random component
which makes them somewhat noisy rotators.  The ultimate causes of this
``timing noise'' are not understood.

Recently it has been shown that the timing behavior of several radio
pulsars is consistent with oscillation between two discrete values of
spin-down rate differing by $\sim 1\%$, which may explain their timing
noise to a significant extent \citep{lhk+10}.  In addition, the pulse
profiles show changes that are correlated with spin-down rate.  This newly
identified connection between torque and radiative properties may be
related to the long-known but also poorly understood ``mode changing''
phenomenon, in which some pulsars abruptly change between two discrete
average radio profiles \citep{bac70a}.

An extreme example of this behavior is apparently provided by \psrb,
with two states lasting for days--weeks in which the spin-down rates
differ by $\sim 50\%$ --- and in the low state, the pulsar ceases to
emit radio pulsations altogether!  The discoverers surmise that (extra)
magnetospheric currents present in the on state are ultimately responsible
for both the radio profile and the extra torque \citep{klo+06}.  It seems
plausible that this could be linked to another long-standing mystery in pulsar
phenomenology, ``nulling'', in which the radio emission from some pulsars
turns off for several rotations before resuming \citep{bac70}.

Extreme intermittency as displayed by \psrb\ is clearly rare, but its true
incidence is not known.  This behavior provides a new probe of pulsar
magnetospheres, and further examples may lead to a better understanding
of some of these phenomena.  Here we report the discovery and initial
study of \psr, which turned off one year after discovery and resumed
pulsations 580 days later.

\section{Observations and Results} \label{sec:obs} 

\subsection{Discovery}

Since the first discovery of radio pulsations from a magnetar
\citep{crh+06}, we had occasionally searched for radio emission from
the magnetar \mag, located at the center of the supernova remnant (SNR)
\snr\ \citep[G27.4+0.0;][]{vg97}.  We had done this at the CSIRO Parkes
telescope in the 20\,cm band (1.4\,GHz), but owing to absurdly strong
radio frequency interference (RFI) caused by the Thuraya-3 satellite,
on Boxing Day 2008 we instead observed in the 10\,cm band.  We observed
the position of the magnetar for 20 minutes at a center frequency of
3078\,MHz, using the analog filterbank/PMDAQ system to sample at 1\,kHz
the total power in each of 288 channels across a bandwidth of 864\,MHz
before recording to disk.  The data were analyzed with standard pulsar
search techniques implemented in PRESTO \citep{ran01}, and a strong
pulsar with period $P=0.912$\,s and dispersion measure $\mbox{DM}=530$\,pc\,cm$^{-3}$
was easily identified.  This is not the magnetar, which has $P=11.78$\,s.

\subsection{Timing, disappearance, and reappearance} \label{sec:timing}

We confirmed the pulsar at the NRAO Green Bank Telescope
(GBT) on 2008 December 31 with a SPIGOT \citep{kel+05}
observation at 2\,GHz and began timing it there on a regular
basis.  After a couple of weeks we switched to using the GUPPI
spectrometer\footnote{https://wikio.nrao.edu/bin/view/CICADA/GUPPiUsersGuide/},
recording data from a bandwidth of 800\,MHz centered on 2\,GHz.
Each observation typically lasted for 5 minutes and we obtained 39 daily
pulse times of arrival between 2009 January 4 and 2010 January 8.

Using TEMPO\footnote{http://tempo.sourceforge.net/} we determined
a phase-connected timing solution for the pulsar, listed in
Table~\ref{tab:parms} (``Solution 1'').  \psr\ is located $4'$ away from
the discovery pointing position, outside the projected extent of the
\snr\ SNR (see Figure~\ref{fig:vlacxo}).  The timing solution contains a
frequency second derivative (where $\nu = 1/P$), nominally significant at
the $6\,\sigma$ level, which whitens the residuals.  Inclusion of $\ddot
\nu$ in the fit changes the values of R.A.\ and $\nu$ by $6\,\sigma$,
decreases the post-fit rms residual from 1.1\,ms to 0.9\,ms,
and presumably reflects timing noise in the pulsar.  While the magnitude
of $\ddot \nu$ suggests a large level of timing noise \citep[see,
e.g.,][]{antt94}, its impact on rotation is still tiny compared to the
effect of $\dot \nu$: during 2009, its overall contribution to pulse
phase, $\ddot \nu t^3/6 = 0.25$, is only 1\% of $\dot \nu t^2/2$.

We carried on studying the pulsar at the GBT during 2010 and into 2011,
but on all 28 observing dates between 2010 January 19 and 2011 July 26
(and on three occasions at Parkes) it was never detected, even after searching the data in period --- then, on
2011 August 11, it reappeared at the GBT as bright as ever!  It seems
that the pulsar turned off for 1.5--1.6\,yr.  The flux density for each
GBT non-detection was $S_2\la0.1$\,mJy, at least 50 times below the
pulsar's average flux when emitting (Table~\ref{tab:parms}).

We do not of course know that the pulsar was off all the time during
those 1.5\,yr, only that in every one of 28 attempts, once every 20 days
on average, for 5 minutes at a time, we never detected it.  Conversely,
we detected it on every one of the 43 days that we observed it spanning
the previous 1.0\,yr.  Both of these facts are suggestive of very long
continuous either on or off states.

However, one atypical observation serves as a
cautionary note: on 2009 December 11 (MJD~55176) the pulsar was detected
as usual in a 300\,s observation at 17.7$^{\rm h}$~UT, but in a second
observation at 19.9$^{\rm h}$ it was not detected.  After confirming that
the equipment was working properly, another 300\,s observation was done
at 20.0$^{\rm h}$ and still the pulsar was not detected.  Finally, at
20.4$^{\rm h}$ the pulsar was detected as normal in a 300\,s observation.
It appears that on this day the pulsar was off for between 10 minutes
(700 rotations) and 2.7\,hr.

The reality is that during 2009 we observed \psr\ for a total of only
4\,hr mostly in $\approx 5$ minute sessions, or one part in 2000 of the
entire year.  During a portion of one daily session out of 40, the pulsar
turned off, likely for a time amounting to a few percent of the total observing
time during the year.  Minding the danger of extrapolating from one
event, we may suppose that the pulsar actually turned off during 2009
in relatively brief episodes amounting cumulatively to $\sim 0.1$--1\%
of the time.  We may also wonder whether during the overwhelmingly off
state in 2010 and into 2011 the pulsar occasionally turned on.

\begin{figure}[t]
\begin{center}
\includegraphics[scale=0.42]{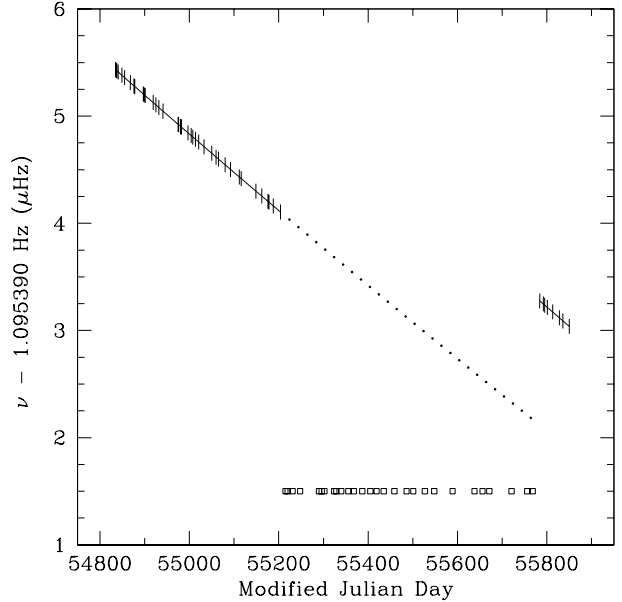}
\end{center}
\caption{\label{fig:timing}
Rotation frequency versus date for \psr.  The first solid line represents
the run of $\nu$ from the phase-connected timing solution obtained
in 2009.  The dotted line is the extrapolation of this trend during
the time when the pulsar was not detected.  Each small square (placed
arbitrarily at a vertical coordinate of 1.5) represents a GBT observing
date during this period.  The small second solid line, much above the
extrapolated trend, is obtained from the timing solution in late 2011,
when the pulsar reappeared.  Vertical tick marks overlaid on the solid
lines represent actual detections.  }
\end{figure}

We have been timing the pulsar since it reappeared, and have detected
it in all nine attempts (once at Parkes).  The $\dot \nu$ of the new
timing solution, measured with 0.5\% precision after 66 days, is the same
as the one measured in 2009 (``Solution 2'' in Table~\ref{tab:parms}).
Extraordinarily, the new rotation frequency is well above that
expected from the extrapolation of the timing solution in 2009 (see
Figure~\ref{fig:timing} and Table~\ref{tab:parms}): the $\dot \nu_{\rm
on}$ measured in 2009 and over the past two months is 2.47 times larger than
the average $\dot \nu_{\rm off} \equiv \Delta \nu/\Delta T$ inferred
for the off state, computed from the difference in rotation frequencies
measured on 2011 August 11 and 2010 January 8 (MJDs~55784 and 55204) from
the respective timing solutions, and $\Delta T=580$\,d.  Considering the
actual observing dates, the off state could have been as many as 27 days
shorter than this, which implies a possible $\dot \nu$ ratio as large
as 2.65.  In principle, the frequency offset in Figure~\ref{fig:timing}
could also be interpreted as a glitch with $\Delta \nu/\nu=10^{-6}$,
but it is rare for such large glitches to be observed in pulsars with
such relatively small $\dot \nu$ \citep{elsk11}, and a glitch has never
resulted in the known disappearance of pulses \citep[although in the young
and high magnetic field PSR~J1119--6127, extra profile components have
been detected following a large glitch;][]{wje11}.  If the mostly
on or mostly off states of \psr\ are contaminated by their complement,
as noted in the previous paragraph, then the true value of the $\dot \nu$
ratio would be slightly increased from the nominal value inferred.  Also,
some of the $\ddot \nu$ fitted in 2009 might be due to contamination
of mostly $\dot \nu_{\rm on}$ by some $\dot \nu_{\rm off}$.  A higher
observing cadence could establish the ``purity'' of the on and off states,
and might also help with the interpretation of $\ddot \nu$.

\subsection{Scattering, pulse profile, and polarimetry} \label{sec:pol}

\begin{figure}[t]
\begin{center}
\includegraphics[scale=0.81]{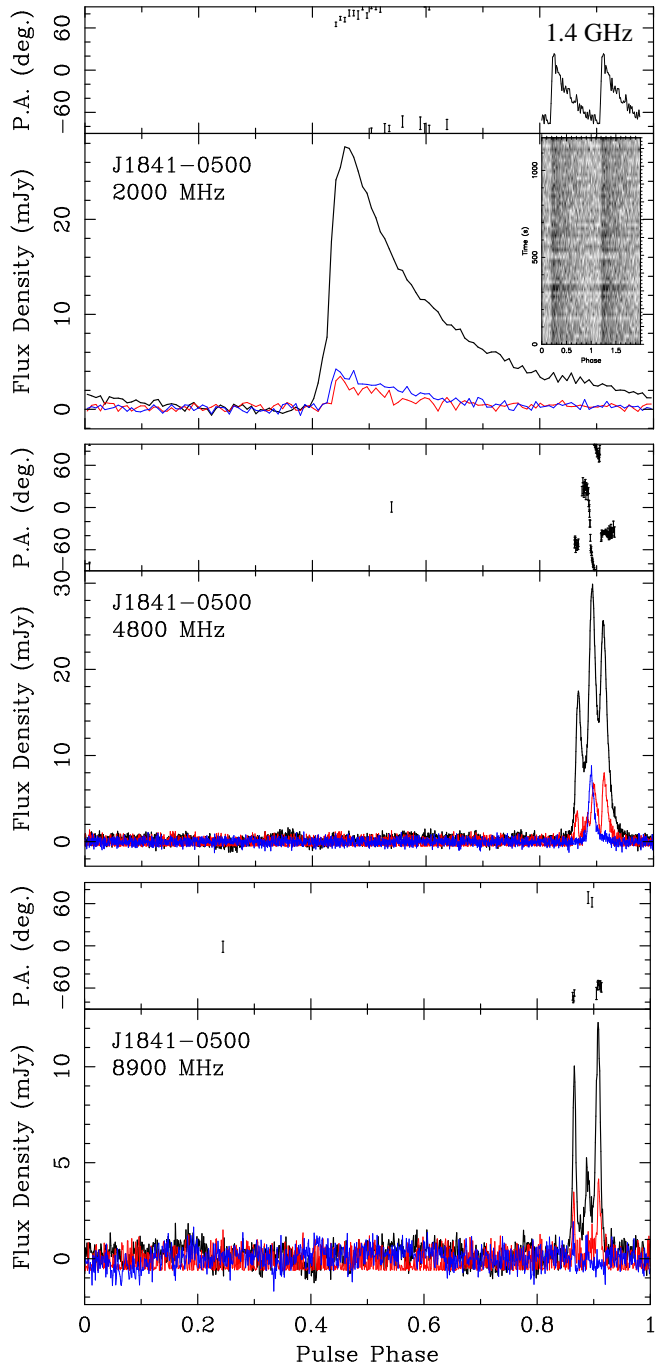}
\end{center}
\caption{\label{fig:pol}
Pulse profiles for \psr\ at four frequencies.  Top: 5 minutes at
2\,GHz from 2011 August displayed with 128 bins.  The inset shows
the only existing 1.4\,GHz detection (20 minutes in 2009 October
at Parkes), with the pulse profile displayed twice, as a function
of time in greyscale and integrated at the top.  Middle: 20 minutes
at 5\,GHz from 2009 September 29 displayed with 2048 bins.  Bottom:
23 minutes at 9\,GHz from 2009 September 29 displayed with 1024 bins.
All polarization data were collected with GUPPI at the GBT.  At each
frequency the black trace corresponds to total intensity, the red one
to linear polarization, and the blue to circular.  In all top subpanels
we display the position angle of linear polarization, corrected to
the pulsar reference frame with $\mbox{RM}=-2993$\,rad\,m$^{-2}$, for
bins where the linear signal-to-noise ratio $>3$.  The 2\,GHz profile
is arbitrarily aligned with respect to the two at higher frequencies,
and its PA values are essentially meaningless because of scattering.
Small wiggles in the off-peak region of the two high-frequency profiles
are due to imperfectly removed RFI.  }
\end{figure}

\begin{figure}[t]
\begin{center}
\includegraphics[scale=0.88,angle=0]{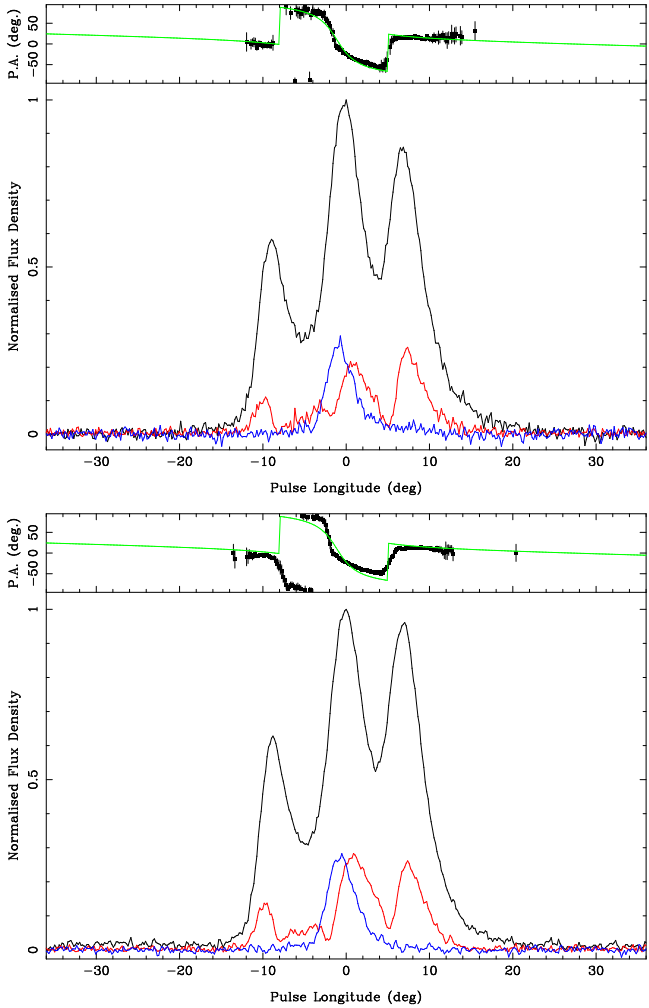}
\end{center}
\caption{\label{fig:rvm}
Rotating vector model fit to 5\,GHz data of \psr.  Top: 2009 profile,
shown in the middle panel of Figure~\ref{fig:pol}, zoomed in on the
emission region, along with RVM curve (green) fitted to position angles
where linear polarization signal-to-noise ratio $>2$ (black).  Bottom:
2011 profile, based on 50 minutes of data.  For comparison, we plot in
green the RVM model obtained for the 2009 data.  In neither case have
the PAs been rotated to the reference frame of the pulsar.  }
\end{figure}

In the discovery observation of \psr, the pulse profile appeared to be
scattered by the ISM, despite the high frequency of 3\,GHz.  This huge
level of scattering is confirmed at $\approx 2$\,GHz (top panel of
Figure~\ref{fig:pol}): our fitted scattering functions yield a $1/e$
scattering timescale of $\tau_1 = (2.29\pm0.02)$\,s, referenced to
a frequency of 1\,GHz --- to our knowledge this is the largest for
any known pulsar \citep[with the possible exception of the magnetar
1E1547.0--5408;][]{crhr07}, and 100 times larger than predicted by
the \citet{cl02} electron density model.  In our fits we assumed an
intrinsic pulse profile made up of three gaussians with the central
one dominant and its ratio compared to the other two extrapolated
from the spectral evolution observed between 5\,GHz and 9\,GHz (see
below and Figure~\ref{fig:pol}).  We also assumed that $\tau_f \propto
f^{-\gamma}$ with $\gamma=4$.  Fitting for the scattering spectral
index results in a somewhat shallower frequency dependence, but the
$\chi^2$ of the fit does not improve by much.  Nevertheless, it appears
that a bit of scattering is still present at 5\,GHz (bottom panel of
Figure~\ref{fig:rvm}), which may support $\gamma<4$, as has been seen
for other pulsars with large DM \citep{bcc+04,lkm+01,lmg+04}.

We had not detected the pulsar at 1.4\,GHz (with 288\,MHz of bandwidth)
on four previous occasions (in 2006 May and 2007 June and December) and
at first assumed that this was due to scattering\footnote{We also did not
detect the pulsar in a subsequent reanalysis of the closest pointing of
the Parkes multibeam survey \citep[][]{mlc+01} to the pulsar position,
$4\farcm5$ away, from 1998 August.  In addition, the pulsar was not
detected by \citet{lkc+11} in a 2006 November GBT search of the nearby
magnetar.}.  In fact the pulsar is detectable at 1.4\,GHz when on, as seen
in the top panel inset of Figure~\ref{fig:pol}.  At this frequency the
profile is substantially scattered by more than the rotation period ---
i.e., some of the flux received at the Earth is not pulsed; however, it is
intrinsically bright enough that the remaining pulsed flux is detectable.

Flux-calibrated observations at 2\,GHz, 5\,GHz, and 9\,GHz (shown in
Figure~\ref{fig:pol}) yield the period-averaged flux densities listed in
Table~\ref{tab:parms}.  The uncertainty at 5\,GHz is from the difference
of two measurements.  We have no evidence for significant flux variation
due to scintillation, although the 2\,GHz observation contains only
about 300 pulses (which may not be enough to stabilize the profile),
and we assume a 20\% uncertainty, as we do at 9\,GHz.  While the nominal
spectrum computed between 5\,GHz and 9\,GHz is steeper than at lower
frequencies, the uncertainties are large enough that we cannot be sure
whether this is significant.  In any case, $\delta \approx -2$ (where
$S_{f} \propto f^{\delta}$), and at 1.4\,GHz the predicted intrinsic
flux density is about 10\,mJy.

Although the profile at 2\,GHz shows little sign of structure because
of the long scattering tail, the 5\,GHz and 9\,GHz profiles (lower
panels of Figure~\ref{fig:pol}) allow us to say a great deal about
the classification and geometry of the pulsar.  The polarimetric
observations were all done with 800\,MHz of bandwidth and analyzed
with PSRCHIVE \citep{hvm04}.  The rotation measure was determined to be
$\mbox{RM}=-3000$\,rad\,m$^{-2}$, which in absolute value is the largest
known for any pulsar.  At 5\,GHz we also detect numerous single pulses,
with peak flux densities up to 200\,mJy, and polarization fractions
ranging over $\sim 0$--100\%.

The average pulse profile has three distinct components with the central
one most prominent at 5\,GHz and significantly less so at 9\,GHz.
The components at 9\,GHz are narrower than at 5\,GHz, possibly as a
result of residual scattering at 5\,GHz.  The outer components (separated
by $0.042\,P$ at both frequencies) do not flank the central component
perfectly symmetrically, with the trailing edge closer.  The linear
polarization is moderate throughout, although characteristically absent on
the leading and trailing wings.  Circular polarization is seen against the
central component but there is no obvious sign reversal.  The polarization
position angle swing is complex with at least two and possibly three
orthogonal mode jumps across the profile (see Figure~\ref{fig:rvm}).

The triple structure with a steep spectrum of the central component
is characteristic of many pulsar profiles \citep{ran93}.  The central
component likely originates near the magnetic pole with the outer
components forming a cone of emission at some reasonable fraction of the
distance to the edge of the polar cap.  The measured full width is some
25\degr.  This is much larger than the $\sim 14\degr$ derived under the
assumption that the emission height is $\sim 300$\,km, typical of older
pulsars \citep{mr02a}.  This then implies that either the emission height
is much larger (greater than 1000\,km) or that the value of $\alpha$
(angle between the magnetic and spin axes) is less than 45\degr.

In the middle panel of Figure~\ref{fig:pol}, the steep position angle of
linear polarization under the main peak hints at a small impact parameter
$\beta$ (closest approach between the observer line of sight and the
magnetic axis).  However, when we look at the profiles and PAs in more
detail (Figure~\ref{fig:rvm}), reality appears much more complicated.
We show in the top panel of Figure~\ref{fig:rvm} a rotating vector
model fit \citep{rc69a} to 5\,GHz data from 2009.  The PAs in the outer
pulse components are apparently offset by 90\degr\ with respect to the
adjoining PAs in the middle component; the fit in the middle component
seems to capture the major trend of PA (and taken at face value implies
$\beta \la 3\degr$) but clearly there are unmodeled details.

The inadequacies of the RVM fit and complicated PA structure of the
profile are seen more clearly in the bottom panel of Figure~\ref{fig:rvm},
where we show 5\,GHz data from 2011, with higher signal-to-noise ratio.
The RVM model shown here is the one obtained from the fit to 2009 data,
and clearly does not fit well the complex progression of PA across the
profile.  \citet{kar09} has shown that orthogonal jumps in PA together
with even modest levels of scattering can distort observed PA swings,
and this may be occurring for \psr\ at 5\,GHz.  In that case, even our
supposition that $\beta$ may be small is not necessarily correct.

The two 5\,GHz profiles differ slightly (e.g., the total intensity
third component is relatively stronger in 2011; there are also slight
differences in polarized flux).  In order to investigate whether these
differences are meaningful, we split the 2011 observation into two
equal halves.  The profiles differ from each other by more than the
noise level, although by a little less than the 2009--2011 difference.
It is conceivable that 1500 rotations are not enough to generate a
stable average pulse profile.  It is also possible that RFI contamination
is responsible for some of these differences (see off-pulse RFI in the
middle panel of Figure~\ref{fig:pol}).  Further observations are required
in order to reach credible conclusions concerning the stability of the
\psr\ profile (the two 5\,GHz and one 9\,GHz observations presented here
are the only ones that exist at frequencies greater than 3\,GHz).

\subsection{The radio and X-ray neighborhood of \psr\ }

\begin{figure*}[t]
\begin{center}
\includegraphics[scale=0.92,angle=0]{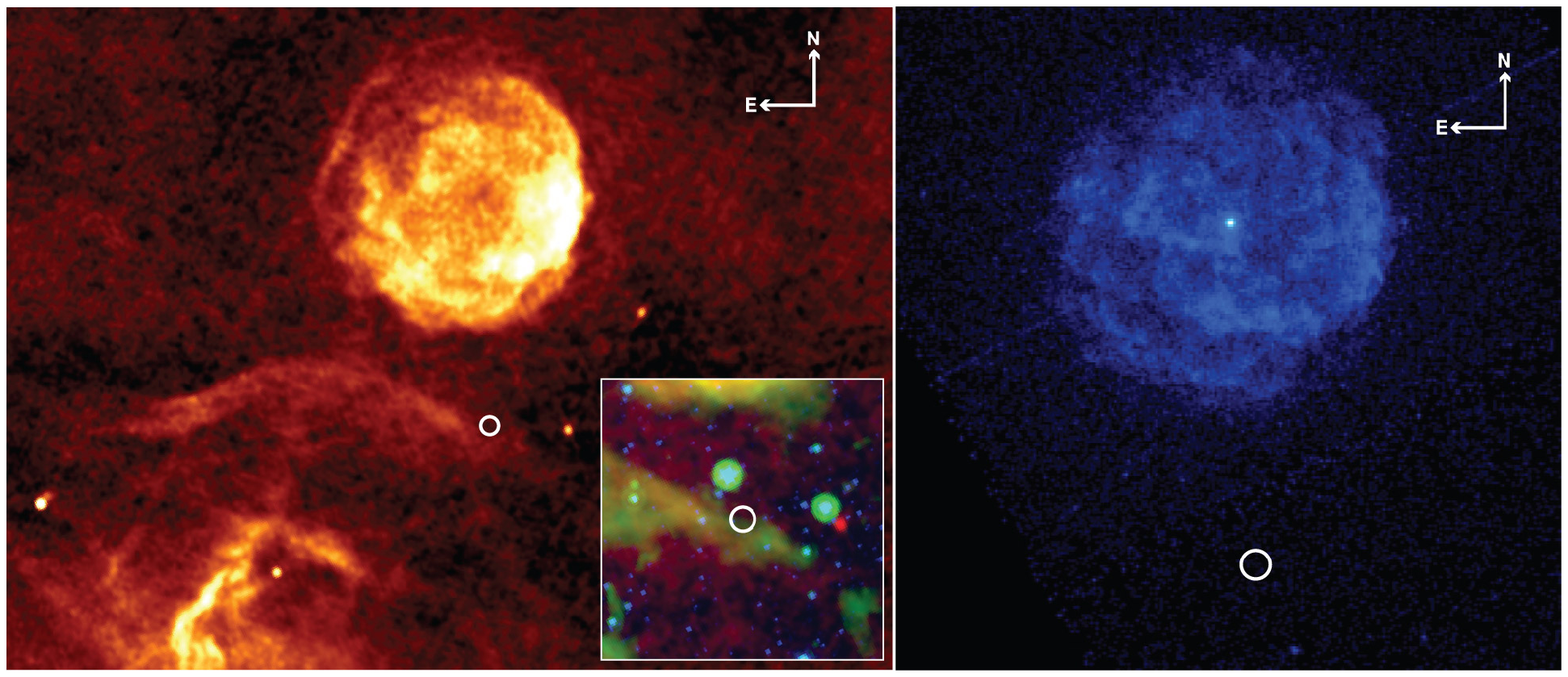}
\end{center}
\caption{\label{fig:vlacxo}
The neighborhood of \psr.  In all panels, the pulsar position (known
with $\approx 1''$ accuracy) is indicated by a circle $10''$ in radius.
Left: MAGPIS radio (1.4\,GHz) image \citep{hbw+06}, $16'\times12'$,
showing the circular SNR~\snr\ and the location of \psr\ to its south.
Inset: $4'\times4'$ multi-wavelength view centered on the pulsar
location, showing MAGPIS 1.4\,GHz in red, MIPSGAL $24\,\mu$m in green,
and GLIMPSE $3.6\,\mu$m in blue.  Right: \cxo\ ACIS-S X-ray (0.5--7\,keV)
exposure-corrected image, $8'\times8'$ with $2''$ resolution, showing
SNR~\snr\ with its central magnetar \mag\ and the position of \psr.
All datasets except for $3.6\,\mu$m have huge dynamic range and are
presented with logarithmic scaling.  }
\end{figure*}

\psr\ is located right on the Galactic plane, at $b = - 0\fdg03$,
and this location has been imaged multiple times at radio wavelengths.
In the left panel of Figure~\ref{fig:vlacxo} we show the most beautiful
existing image of its surroundings.  This is extracted from the MAGPIS
survey \citep{hbw+06} at a wavelength of 20\,cm, and the image has an
approximate angular resolution of $6''$ and rms sensitivity of 0.3\,mJy.
\psr\ (with position indicated by a circle) is not detected, with
a $3\,\sigma$ upper limit of 1\,mJy (the point sources to the west
and northwest of the pulsar have flux densities of 6\,mJy and 5\,mJy,
respectively).  At 2\,GHz the pulsar flux density is $S_2 = 5$\,mJy, and
we estimate (Section~\ref{sec:pol}) that $S_{1.4} \approx 10$\,mJy ---
when the pulsar is turned on!  Evidently, when the multi-epoch VLA and
Effelsberg observations for this image were done, the pulsar was not on.

The SNR~\snr\ and its magnetar have been observed with the \cxo\
X-ray Observatory, and the new pulsar falls within the field of
view.  We have reprocessed two observations using the latest CIAO
pipeline\footnote{http://cxc.harvard.edu/ciao/}, from which we determine
an upper limit for X-ray emission from \psr.

In a 29\,ks ACIS-I observation (ObsID~729, done in 2000 July), only
one photon is detected in the 0.5--7\,keV range within a circle of
radius $2''$ centered on the pulsar.  The background level is high
due to the nearby SNR, and the source falls near a chip boundary.
The exposure-corrected upper limit on the 0.5--7\,keV count rate is
$1.2\times10^{-4}$\,s$^{-1}$ at the 90\% confidence level.  A slightly
better exposure-corrected 90\% upper limit of $9.4\times10^{-5}$\,s$^{-1}$
is derived from a 25\,ks ACIS-S observation (ObsID~6732, done in 2006
July), in which no counts are detected within the same aperture.

Assuming a putative power-law spectrum with photon index $\Gamma=1.5$
absorbed by a column with $N_H = 1.6\times10^{22}$\,cm$^{-2}$
(corresponding to one free electron for every 10 neutral hydrogen atoms
along the line of sight), this implies an unabsorbed flux of $f_X <
2\times10^{-15}$\,erg\,cm$^{-2}$\,s$^{-1}$, or luminosity $L_X < 10^{30}
(d/7\,{\rm kpc})^2$\,erg\,s$^{-1}$.  For a distance of 7\,kpc, as inferred
from the DM and the \citet{cl02} model, $L_X/\dot E < 6\times10^{-4}$.
For SNR~\snr, at $7.5\,\mbox{kpc} < d < 9.4$\,kpc \citep{tl08}, $N_H =
(2-3)\times10^{22}$\,cm$^{-2}$ \citep{vg97}; the huge amount of scattering
for \psr\ (Section~\ref{sec:pol}) also suggests that there might be more
X-ray absorption or scattering than assumed, so that a more realistic
limit might be $L_X/\dot E < 10^{-3}$ or even higher.  At such levels,
the non-detection of \psr\ in X-rays is not surprising \citep[see,
e.g.,][]{pccm02}.

\section{Discussion} \label{sec:disc} 

\psr\ is an extremely intermittent pulsar, the only one besides \psrb\
to have been reported on in detail.  Some differences are that the off
state of \psr\ is $\sim 20\times$ longer than for \psrb, the ratio of
on-to-off $\dot \nu$ is 2.5 rather than 1.5, and at this point we still
do not know whether the on--off cycles in \psr\ occur quasi-periodically,
since we have only observed one off interval surrounded by two on phases,
the first of which lasted for at least 1\,yr.  (The five non-detections
in 2006--2007 are consistent with another off period lasting also
for $\sim 600$ days; see Section~\ref{sec:pol}.)  Although published
details are limited, a third such pulsar is known: \psrc\ \citep{lfl+06}
was observed to turn off for about 2\,yr between two 1\,yr on phases,
with a ratio of on-to-off $\dot \nu$ of 1.7 \citep{kra08,lyn09}.

We do not know what causes these (or any) pulsars to turn off, or back on.
But any plausible magnetospheric model should attempt to account for
the observed on-to-off ratios of $\dot \nu$, which now span 1.5--2.5 for
three pulsars.  Models of pulsar magnetospheres range between those that
contain no plasma, where $\dot \nu \propto 2 \sin^2\alpha/3$ \citep[e.g.,][]{mt77}, and ideal
MHD plasma-filled descriptions where $\dot \nu \propto 1 + \sin^2\alpha$ \citep{spi06,kc09}.
None of these can explain actual pulsar emission: the vacuum models
have no charged particles that can be accelerated and radiate, while the
``force-free'' models have $\vec{E} \cdot \vec{B}=0$ and cannot accelerate
the charged particles to radiate.  On the plus side, quantitative
solutions exist for these limiting cases for arbitrary inclinations
$\alpha$ \citep[see, e.g.,][]{spi06}.  In the context of these models
it may be tempting to associate the pulsar on states with a force-free
magnetosphere and the pulsar off states with a vacuum magnetosphere,
but the $\dot \nu$ ratios for those models are always $>3$, not matched
by the observations.  Recently, \citet{lst11} and \citet{kkhc11} have
computed resistive solutions for pulsar magnetospheres that in significant
respects lie between these extrema.  Associating the on state with the
force-free magnetosphere and the off state with a resistive configuration
(i.e., with a suppression of the conduction current, for unknown reasons),
these finite conductivity models can produce the observed $\dot \nu$
ratios for any value of $\alpha$, including intermediate ones (which we
expect a priori, given that only three of these intermittent pulsars are
known; for smaller $\alpha$, the $\dot \nu$ ratios should be larger,
all else being equal).  It seems encouraging that these advances in
the treatment of pulsar magnetospheres provide a framework that may
help to explain the unexpected observations of two-state $\dot \nu$
in PSRs~B1931+24, J1832+0029, and J1841--0500.

Pulsar intermittency is observed with a huge variety of on and off
timescales.  Rotating radio transients (RRATs) are sporadic emitters
of isolated radio pulses \citep{mll+06}, and while much remains unclear
about RRATs, some of them may represent simply a more extreme form of the
classical nulling phenomenon \cite[see][]{bb10}, with nulling fractions
of up to $>99\%$.  Where do \psr\ and its two identified cousins fit in
this scheme?  Their overall nulling fractions (about 20--40\%) are not
particularly large, but the average time off between on states {\em is}
exceptional: for nulling pulsars and RRATs, this ranges smoothly between
seconds and $\approx 1$\,hr, but it is $\approx 1$ month for \psrb,
and $\approx 1$\,yr for \psr.  Based in part on these observations,
\citet{bbj+11} argue that the intermittency of \psrb\ represents a
distinct phenomenon to nulling/RRATing, and this conclusion applies even
more so to \psr.  What type of pulsar can display extreme intermittency
like PSRs~B1931+24, J1832+0029, and J1841--0500?  RRATs, on average, may
occupy a special location in the pulsar $P$--$\dot P$ diagram, with longer
periods and larger inferred surface magnetic field strengths than the
bulk of the population \citep{mlk+09}.  Magnetars can be transient radio
emitters \citep[e.g.,][]{crh+06}, but they certainly occupy a distinct
quadrant of the $P$--$\dot P$ diagram.  But PSRs~B1931+24, J1832+0029,
and J1841--0500 as a group occupy an unremarkable location in $P$--$\dot
P$, and apart from their intermittency and associated torque changes
there does not appear to be anything exceptional about them.  Can other
apparently ordinary pulsars display this behavior at some point during
their $\sim 10^7$\,yr radio lifetimes, perhaps with even longer off (and
on) timescales?  Maybe some pulsar that has been observed steadily for
decades will one of these years disappear, not to return in our lifetimes,
but returning eventually after centuries dormant.

Quite apart from its unusual radiative and rotational
properties, \psr\ is somewhat unusual in being located
only $4'$ from another (very young) neutron star, the
magnetar \mag\ within the SNR~\snr.  In the ATNF pulsar catalog
\citep{mhth05}\footnote{http://www.atnf.csiro.au/research/pulsar/psrcat/},
there are 11 neutron star pairs listed with separations $<4'$, among
1794 pulsars in the Galactic disk.  The probability of finding such a
pair by chance is therefore of order 1\%.  The distance estimates that we have
for both of these objects are compatible, at approximately 8.5\,kpc.
The question therefore arises of whether this is merely a coincidence.
At that distance, the angular separation corresponds to 10\,pc.  If both
progenitor stars had been in a binary system that disrupted upon the
second supernova explosion (SNe), then for an assumed magnetar/SNR age of
10\,kyr, the implied transverse velocity is nearly 1000\,km\,s$^{-1}$,
which is large but not impossibly so.  However, one might then have
expected evidence for the passage of a high-velocity neutron star through
the SNR shell \citep[see, e.g.,][]{cng+09}.  In addition, the SNR is
probably much younger \cite[$\approx 1000$\,yr according to][]{tl08}.
If on the other hand the putative binary had disrupted upon the first SNe,
the implied transverse velocity is much smaller, $\sim 10$\,km\,s$^{-1}$.
Even if \psr\ and \mag\ are not directly associated, it is plausible
that they might both have been born in the same young stellar cluster.

The remaining puzzle concerning \psr\ is its huge RM and amount of
scattering.  At a distance of 7\,kpc, the pulsar is located on the inside
edge of the Scutum spiral arm, where RMs are predominantly positive
with $|\mbox{RM}|<500$\,rad\,m$^{-2}$, reflecting the counterclockwise
large-scale magnetic field of the Crux-Scutum arm \citep{hml+06}.
For \psr, the $\mbox{RM}=-3000$\,rad\,m$^{-2}$ must be largely due to
one or more discrete intervening sources along the line of sight, for
instance an SNR.  The anomalously large level of scattering is presumably
also caused by discrete sources, for instance an H{\sc ii} region or SNR.
In the left panel of Figure~\ref{fig:vlacxo} we see that the pulsar is
located in projection extremely close to the boundary of an arc that may
surround a larger structure to the south.  It is not clear whether this
``wisp'' is thermal or non-thermal in nature.  The radio filament at
20\,cm exactly overlaps a $24\,\mu$m filament detected in the MIPSGAL {\em
Spitzer} infrared survey\footnote{See http://mipsgal.ipac.caltech.edu.
Multi-wavelength views of this busy region of the Galactic plane
can be obtained at http://third.ucllnl.org/cgi-bin/colorcutout.}
(see the inset in Figure~\ref{fig:vlacxo}).  However, there
is no corresponding emission detected in the GLIMPSE survey at
3--8\,$\mu$m\footnote{http://www.astro.wisc.edu/sirtf/}, suggesting that
the dust in this region is fairly cold and likely not part of an H{\sc ii}
region (in contrast, the bright filament to the southeast of the pulsar,
aligned along a southeast--northwest direction, is clearly thermal).
Perhaps this is an old SNR shell filament that could account for the
scattering and the high line-of-sight magnetic field for \psr, if the
pulsar lies behind it.

Much of the Galactic plane has been surveyed for pulsars with good
sensitivity, but for the most part it has not been searched {\em
repeatedly}.  The detection of intermittent pulsars would benefit
from long-term wide-field imaging surveys for slow radio transients.
For pulsars that are as, or more, scattered than \psr, traditional surveys
require a high frequency, with the drawbacks of small telescope beam
size and small pulsar flux \citep[see][]{bjl+11}.  Imaging surveys might
prove to be a more efficient method to discover such pulsars.

\acknowledgements

We are grateful to David Helfand, Anatoly Spitkovsky, Duncan Lorimer,
Joanna Rankin, and Alice Harding for useful discussions.  We thank Marta
Burgay for help identifying Parkes archival data.  FC completed much of
this work at the Turramurra Operations Centre (TOC) and thanks all the
staff, A.\ Brown, A.\ `Gus' Lord, H.\ Reynolds, and J.\ Reynolds, for
tremendous hospitality.  SC acknowledges support from the National Science
Foundation (NSF) through grant AST-1008213.  The Parkes Observatory is
part of the Australia Telescope, which is funded by the Commonwealth
of Australia for operation as a National Facility managed by CSIRO.
The National Radio Astronomy Observatory is a facility of the National
Science Foundation operated under cooperative agreement by Associated
Universities, Inc.

{\em Facilities:}  \facility{Parkes (PMDAQ)}, \facility{GBT (GUPPI)}

\begin{deluxetable*}{ll}
\tablewidth{0.60\linewidth}
\tablecaption{\label{tab:parms} Measured and Derived Parameters for
PSR~J1841--0500 }
\tablecolumns{2}
\tablehead{
\colhead{Parameter} &
\colhead{Value}
}
\startdata
Data span (MJD)\dotfill          & 54835--55204 (Solution 1)                  \\
Right ascension, R.A. (J2000.0)\dotfill & $18^{\rm h}41^{\rm m}18\fs14(5)$    \\
Declination, decl. (J2000.0)\dotfill    & $-05\arcdeg00'19\farcs5(8)$         \\
Rotation frequency, $\nu$ ($s^{-1}$)\dotfill & 1.0953947642(3)                \\
Frequency derivative, $\dot \nu$ ($s^{-2}$)\dotfill 
                                        & $-4.165(1)\times10^{-14}$           \\
Frequency second derivative, $\ddot \nu$ ($s^{-3}$)\dotfill 
                                        & $4.8(8)\times10^{-23}$              \\
Epoch of frequency (MJD)\dotfill        & 55018.0                             \\
\hline
Data span (MJD)\dotfill          & 55784--55850 (Solution 2)\tablenotemark{a} \\
Rotation frequency, $\nu$ ($s^{-1}$)\dotfill & 1.0953931969(2)                \\
Frequency derivative, $\dot \nu$ ($s^{-2}$)\dotfill 
                                        & $-4.17(2)\times10^{-14}$            \\
Epoch of frequency (MJD)\dotfill        & 55800.0                             \\
\hline
Dispersion measure, DM (pc\,cm$^{-3}$)\dotfill & $532\pm1$\tablenotemark{b}   \\
Rotation measure, RM (rad\,m$^{-2}$)\dotfill            & $-2993\pm50$        \\
Flux density at 2\,GHz, $S_2$ (mJy)\dotfill             & $5.4\pm1.1$         \\
Flux density at 5\,GHz, $S_5$ (mJy)\dotfill             & $1.1\pm0.1$         \\
Flux density at 9\,GHz, $S_9$ (mJy)\dotfill             & $0.23\pm0.05$       \\
\hline
Galactic longitude, $l$ (deg)\dotfill                   &  27.32              \\
Galactic latitude, $b$ (deg)\dotfill                    & --0.03              \\
DM-derived distance, $d$ (kpc)\dotfill                  & 7                   \\
Characteristic age, $\tau_c$ (yr)\tablenotemark{c}\dotfill 
                                           & $\{0.4,1.0\}\times10^6$          \\
Spin-down luminosity, $\dot E$ (${\rm erg\,s^{-1}}$)\tablenotemark{c}\dotfill
                                           & $\{1.8,0.7\}\times10^{33}$       \\
Surface dipole magnetic field strength (Gauss)\tablenotemark{c}\dotfill  
                                           & $\{5.7,3.6\}\times10^{12}$ \\[-5pt]
\enddata
\tablecomments{Numbers in parentheses represent the nominal $1\,\sigma$
TEMPO timing uncertainties on the last digits quoted.  }
\tablenotetext{a}{For this solution, the celestial coordinates were held
fixed at the value obtained in Solution 1.  }
\tablenotetext{b}{This is a scattering-corrected DM.}
\tablenotetext{c}{These derived quantities are computed using the value
of $\dot \nu$ in the on and off state, respectively.  }
\end{deluxetable*}

\end{document}